\documentclass[aps,rmp,preprint,groupedaddress,floatfix]{revtex4}
\usepackage{epsfig}
\usepackage{amsmath}
\usepackage{genetics}

\begin{document}

\newcommand{\mon}{\begin{displaymath}}
\newcommand{\moff}{\end{displaymath}}
\renewcommand{\b}[1]{\mbox{\boldmath ${#1}$}}
\newcommand{\pd}[2]{\frac{\partial {#1}}{\partial {#2}}}
\newcommand{\od}[2]{\frac{d {#1}}{d {#2}}}
\newcommand{\inti}{\int_{-\infty}^{\infty}}
\newcommand{\eon}{\begin{equation}}
\newcommand{\eoff}{\end{equation}}
\newcommand{\e}[1]{\times 10^{#1}}
\newcommand{\chem}[2]{{}^{#2} \mathrm{#1}}
\renewcommand{\sb}{s}
\newcommand{\s}{s}
\newcommand{\eq}[1]{Eq. (\ref{#1})}
\newcommand{\ev}[1]{\langle #1 \rangle}
\newcommand{\mat}[1]{\bf{\mathcal{#1}}}
\newcommand{\fig}[1]{Fig. \ref{#1}}
\newcommand{\degrees}{\,^{\circ}\mathrm{C}}
\renewcommand{\log}{\ln}
\renewcommand{\sec}[1]{section \ref{#1}}
\newcommand{\pr}[1]{P_{#1}}

\newcommand{\thetaseq}{\theta_l}
\newcommand{\thetasite}{\theta_s}
\renewcommand{\u}{u}
\renewcommand{\L}{\zeta}
\newcommand{\like}{L}

\renewcommand{\baselinestretch}{1.0}

\title{Detecting Directional Selection from the Polymorphism Frequency Spectrum}
\author{Michael M. Desai$^{1}$}
\author{Joshua B. Plotkin$^{2}$}

\affiliation{${}^1$Lewis-Sigler Institute for Integrative Genomics
\\Princeton Univ., Princeton, NJ 08544\\ ${}^2$Department of
Biology\\University of Pennsylvania, Philadelphia, PA 19104}
\date{\today}

\begin{abstract}

The distribution of genetic polymorphisms in a population contains
information about the mutation rate and the strength of natural
selection at a locus.  Here, we show that the Poisson Random Field (PRF)
method of population-genetic inference suffers from systematic biases
that tend to underestimate selection pressures and mutation rates, and
that erroneously infer positive selection. These problems arise from the
infinite-sites approximation inherent in the PRF method. We introduce
three new inference techniques that correct these problems.  We present
a finite-site modification of the PRF method, as well as two new methods
for inferring selection pressures and mutation rates based on diffusion
models. Our methods can be used to infer not only a ``weighted average"
of selection pressures acting on a gene sequence, but also the
distribution of selection pressures across sites. We evaluate the
accuracy of our methods, as well that of the original PRF approach, by
comparison with Wright-Fisher simulations.  \end{abstract}

\maketitle

\section{Introduction}

The mutation rate and selection pressures operating on genes are of
central importance in shaping their evolution.  The number and frequency
distribution of genetic polymorphisms within a population carry
information about these fundamental processes.  Polymorphisms at higher
frequencies reflect weaker selective pressures (or positive selection),
and vice versa.  Similarly, a larger number of polymorphisms  indicates
a higher mutation rate.  Thus we can use the polymorphism frequency
spectrum observed in genetic sequences sampled from a population in
order to infer the mutation rate and the strength and direction of
selection acting on the sequence.

This intuition can be formalized into a rigorous method for estimating
selection pressures and mutation rates by calculating the likelihood of
sampled polymorphism data as a function of these parameters.  The
Poisson Random Field (PRF) model provides an important and widely-used
method of doing so. The PRF model assumes a panmictic population of
constant size, free recombination, infinite sites, no dominance or
epistasis, and equal selection pressures at all sites.  Under these
assumptions, \citet{sawyerhartl92} showed that the distribution of
frequencies of mutant lineages in a population forms a Poisson random
field whose properties depend on the selection pressure and the mutation
rate. \citet{hartlsawyer94} and \citet{bustamante01} developed a maximum
likelihood method of estimating these parameters from data on the
polymorphism frequency spectrum.  This method has been widely used to
study, for example, purifying selection on synonymous
\cite{hartlsawyer94, akashishaeffer97, akashi99} and nonsynonymous
\cite{akashi99,hartlsawyer94} variation, and the evolution of base
composition \cite{lercherhurst02, galtier06}.

Closely related to these analyses of polymorphism data are methods that
calculate, based on the PRF model, the ratio of the expected number of
polymorphisms within species to divergence between species for
synonymous and nonsynonymous sites (using the idea behind the
McDonald-Kreitman test \cite{mcdonaldkreitman91}). These methods discard
some of the available data, as they depend only on the number of
polymorphisms and not their full frequency spectrum. However, they are
also less sensitive to assumptions \cite{loewe06, sawyerhartl92}. Such
methods have been applied to estimate selection pressures on synonymous
variation \cite{akashi95}, on nonsynonymous mutations in mitochondrial
genomes \cite{nachman98, randkann98, weinreichrand00}, and on
nonsynonymous variation in a variety of nuclear genomes
\cite{bustamante02, bartolome05, sawyerhartl03}, including humans
\cite{bustamante05}.

Much recent theoretical work has focused on relaxing various assumptions
of the original PRF method.  These include allowing for dominance
\cite{williamson04}, population subdivision \cite{wakeley03}, changing
population size \cite{williamson05}, and linkage between sites
\cite{zhubustamante05}. Several methods for studying the properties of
the distribution of selection pressures across sites based on the PRF
model have also been developed, both using the polymorphism frequency
spectrum \cite{bustamantehartl03}, and using the ratio of polymorphism
to divergence \cite{loewe06, sawyerhartl03, bustamantehartl03,
piganeaueyrewalker03}.

All of the methods summarized above fall within the PRF framework,
and therefore depend on the infinite-sites approximation.  Rather
than calculating the evolutionary dynamics at each site, these
methods consider the overall steady-state distribution of mutant
lineage frequencies across all sites.  The PRF method is applied to
data by assuming that each lineage segregates at a different site.
The infinite-sites assumption is made for purely technical, as
opposed to biological, reasons.  In this paper, we show that in
biologically relevant parameter regimes the infinite-sites
assumption causes the PRF method to underestimate selection
pressures and mutation rates. This problem arises both for
inferences based on the polymorphism frequency spectrum and for
inferences based on the ratio of within-species polymorphism to
between-species divergence, but in this paper we focus exclusively
on the former.  As we demonstrate below, the PRF method often
underestimates the selection pressure and the mutation rate by as
much as an order of magnitude. In addition, and perhaps of greater
concern, the PRF method frequently infers that a gene is under
strong positive selection when in fact the gene is experiencing
weak negative selection (Fig. 1).

In this paper, we present three methods to correct the systematic biases
of the PRF method, each with their own advantages and drawbacks.  Rather
than study mutant lineages across a sequence, our methods all focus on
explicit models of the evolutionary dynamics at individual sites.  We
first present a modification of the PRF method that calculates the
frequency distribution of mutant lineages at each site, rather than
across the whole sequence.  We next present two new methods based on
well-known diffusion equations in place of the PRF framework.  All three
of our methods allow us to estimate the selection pressure and the
mutation rate from data on the polymorphism frequency spectrum. In
addition, these methods also allow us to infer the distribution of
selection pressures across sites.  In order to assess the accuracy of
these methods, we generate polymorphism data from simulated
Wright-Fisher populations with known selection pressures and mutation
rates.  By comparing inferences drawn from these simulated data sets, we
demonstrate that our methods correct the biases inherent in the original
PRF approach.

\section{The Poisson Random Field Model of Polymorphisms}

We begin by outlining the Poisson Random Field (PRF) model of the
site-frequency spectrum developed by Sawyer and Hartl
\cite{sawyerhartl92, hartlsawyer94}. This model assumes that
mutations occur in a population of effective size $N$ at a Poisson
rate $N u$, where $u$ is the \emph{per-sequence} mutation rate, and
are all subject to selection of strength $s$.  The fate of each
mutant lineage is modeled by a diffusion approximation to the
processes of selection and drift.  When a new mutant lineage enters
the population, it is assumed to arise at a site that has not
previously experienced any mutations (the infinite-sites
assumption).  Each mutant lineage is assumed to be independent of
all others (the free-recombination assumption). Since new mutations
are continuously arising, and older mutant lineages are fixing or
dying out, there is a steady-state distribution of the frequencies
of segregating (i.e. non-fixed and non-extinct) mutant lineages.
The expected number of segregating sites increases linearly with the
mutation rate, but $u$ does not affect the shape of the steady-state
distribution of segregating mutant frequencies.

Extending earlier work by \citet{moran59} and \citet{wright38},
\citet{sawyerhartl92} calculated this steady state distribution of
lineage frequencies. They found that the number of lineages with
frequency between $x$ and $x+dx$ is Poisson distributed with mean
$f(x)dx$, where \eon f(x) = \thetaseq \frac{1 - e^{-2 \gamma
(1-x)}}{1 - e^{-2 \gamma}} \frac{1}{x(1-x)}. \label{prfsteadystate}
\eoff Here $\gamma \equiv N s$ is a measure of the strength of
selection on the mutant lineages and $\thetaseq \equiv 2 N u$ is
twice the population per-sequence mutation rate.  The function
$f(x)$ is referred to as the mean of a Poisson Random Field. In
other words, the number of mutant lineages with frequency between
$x_1$ and $x_2$ is a Poisson random variable with mean
$\int_{x_1}^{x_2} f(x) dx$. In addition, the number of mutant
lineages with frequency in $[x_1,x_2]$ is independent (as a random
variable) from the number of mutant lineages with frequency in
$[y_1,y_2]$, provided these intervals do not intersect.  Note that
$f(x)$ is not integrable at $0$ or $1$. This divergence arises
because the steady state is due to a balance between new mutations
constantly occurring and older lineages fixing or going extinct.
Thus there is no finite, steady-state expression for the number of
lineages that have fixed or gone extinct.

\citet{hartlsawyer94} and \citet{bustamante01} used
\eq{prfsteadystate} as the basis for maximum-likelihood (ML)
estimation of the mutation rate $\thetaseq$ and selection pressure
$\gamma$ from polymorphism data. They imagined sampling and
sequencing $n$ individuals from a population with this steady state
distribution of segregating mutant lineages.  They made the
infinite-sites assumption that all mutant lineages occur at
different sites, consistent with the earlier assumption that each
lineage is independent (used to derive \eq{prfsteadystate}).  If a
given site has a mutant lineage at frequency $x$ in the entire
population, then the probability that a sample of $n$ individuals
will contain $i$ mutant nucleotides and $n-i$ ancestral nucleotides
at the site is ${n \choose i} x^i (1-x)^{n-i}$.  That is, for each
mutant lineage with frequency $x$, there is a probability ${n
\choose i} x^i (1-x)^i$ of finding a corresponding site with $i$
mutant nucleotides.  Since the number of mutant lineages at
frequency $x$ in the population is Poisson distributed with mean
$f(x)dx$, the number of sampled sites containing $i$ mutant
nucleotides (we refer to these as $i$-fold mutant sites) is Poisson
distributed with mean \eon F(i) = \thetaseq \int_0^1 \frac{1-e^{-2
\gamma(1-x)}}{1-e^{-2 \gamma}} \frac{1}{x(1-x)} {n \choose i} x^i
(1-x)^{n-i} dx. \eoff

This equation leads immediately to a maximum likelihood procedure for
estimating $\gamma$ and $\thetaseq$ \cite{bustamante01}.  A set of
sequences from $n$ sampled individuals within a population will contain
some number, $y_i$, of $i$-fold mutant sites for $0< i< n$. The set
of values $y_1$, $y_2$, $\ldots$ $y_{n-1}$ is called the site-frequency
spectrum of the observed data.  The probability of a spectrum $\{y_i\}$,
given $\gamma$ and $\thetaseq$, is \eon L_u(\thetaseq, \gamma) =
\prod_{i = 1}^{n-1} e^{-F(i; \gamma, \thetaseq)} \frac{\left[ F(i;
\gamma, \thetaseq) \right]^{y_i}}{y_i!}.  \eoff For an observed spectrum
{$y_i$} in a particular data set, we can maximize this likelihood over
$\thetaseq$ and $\gamma$ to estimate the mutation rate and selection
pressure.

The likelihood expression above assumes we know which nucleotide is
ancestral and which nucleotide is the mutant at each polymorphic
site. We refer to this situation as the unfolded case. When we do
not have this information, we cannot distinguish between an
$i$-fold mutant site and an $(n-i)$-fold mutant site.  In this
case, a data set will contain some number, $y_i$, $i$-fold and/or
$(n-i)$-fold mutant sites, where $i$ runs between $1$ and the
largest integer less than or equal to $n/2$. We refer to this as
the folded case. In this situation, the likelihood of a particular
data set $L_f(\thetaseq, \gamma)$ is given by the same expression
as in $L_u$, but with the product running from $1$ to $n/2$ and
with $F(i)$ replaced by $F(i) + F(n-i)$ (except if $i = n/2$).

\subsection{Shortcomings of the Poisson Random Field model}

The PRF model makes two key assumptions: that each site is independent
of all the others, and that two mutant lineages never segregate at the
same site.  The former assumption is equivalent to assuming free
recombination between all sites which segregate concurrently. This
assumption may be violated in many populations, particularly when the
method is applied to estimate selective pressures on short stretches of
DNA (e.g. a single gene) that are linked over long timescales.  Ideally,
we would like a theory that allows us to infer the strength of selection
acting on a large number of sites with an arbitrary degree of linkage,
but no such theory yet exists. Nevertheless, a free recombination model
is useful as a null model and a limiting case, and it can be used to
compare with more complicated possibilities.  Such a model may also be
used to test whether recombination is necessary to explain data from a
particular population. Furthermore, whenever selection pressures are
weak, sites segregate over long timescales and so recombination may be
frequent enough that even intragenic sites are unlinked over these
timescales. Finally, it is important to note that linkage between
segregating sites will not bias estimates of the selection pressure,
provided it is equal across sites, but rather increase the variance in
such estimates \cite{bustamante01,akashishaeffer97}.

The second assumption of the PRF, that there are an infinite number
of sites, is more problematic. This problem is most apparent when
considering how the PRF method treats ``multiply polymorphic''
sites --- those that exhibit more than two types of segregating
nucleotides. We refer to the configuration of a particular site as
$(a,b,c,d)$, where $a$, $b$, $c$, and $d$ are the numbers of
sampled sequences which exhibit each of the four nucleotides. When
we have unfolded data, $a$ is the frequency of the ancestral
nucleotide and $b$, $c$, and $d$ are the frequencies of the three
possible mutant nucleotides, in order of decreasing frequency. When
we have folded data, $a$, $b$, $c$, and $d$ are the frequencies of
all four possible nucleotides, again in order of decreasing
frequency. In the original PRF analysis, a site with a $(12,1,1,0)$
configuration, for example, is treated identically to a site with a
$(12,2,0,0)$ configuration.  Such a treatment is incorrect: the
former configuration can only arise from two low-frequency mutant
lineages, whereas the latter configuration could be caused by a
single high-frequency lineage. Yet the PRF analysis excludes the
first possibility, and treats both configurations as if they were
all $(12,2,0,0)$ sites \cite{hartlsawyer94, bustamante01}.
Similarly, the PRF method treats $(10,2,2,0)$, $(10,3,1,0)$, and
$(10,2,1,1)$ sites as if they were in a $(10,4,0,0)$ configuration,
etc.

The infinite sites approximation also affects sites that are not
multiply polymorphic.  The essential problem is that the sampling
mechanism used to calculate $F(i)$ under the PRF method implicitly
assumes that each mutant lineage occurs at a different site. The
number of $i$-fold mutant sites is assumed to equal the number of
$i$-fold mutant lineages sampled. Yet if multiple lineages are
segregating at the same site, an $i$-fold and a $k$-fold mutant
lineage sampled at the same site can lead to an apparently
$(i+k)$-fold mutant site, if the two lineages happen to be
mutations to the same nucleotide.  In other words, a $(12,2,0,0)$
site could reflect two low-frequency mutant lineages or one
higher-frequency one, but the PRF incorrectly assumes that only the
latter is possible.

The infinite-sites assumption is problematic whenever there are
multiple mutations segregating at a site, even if they are at low
frequency. Since a mutant lineage will survive on average $O(\log
\left[ 1/|s| \right])$ generations before fixing or going extinct,
and mutations arise at rate $N \mu$ per site, the infinite-sites
approximation will be valid only when $N \mu \log \left[ 1/|s|\ll
\right] 1$. This condition is often violated in real populations.
In fact, several estimates of mutation rates and selective
pressures based on the PRF method violate this condition
\cite{hartlsawyer94}.  We can also see explicitly that this
condition is violated whenever a data set includes multiply
polymorphic sites. Polymorphisms of this type are indeed observed
in data analyzed by PRF method \cite{hartlsawyer94}.

Unlike the assumption of free recombination, the assumption of
infinite sites is purely technical.  In other words, we do not
aspire to use the PRF method as a null model for exploring whether
or not real systems violate the infinite sites assumption.  This
assumption is highly problematic because it induces systematic
biases in the estimates of selection and mutation obtained by the
PRF method: the method always treats an $i$-fold sampled site as
having arisen from a single mutant lineage that has reached
frequency $i$ in the sample, ignoring the possibility of multiple
mutant lineages at lower frequencies which sum to $i$. By
disregarding the latter possibilities, the PRF method
systematically underestimates the mutation rate $u$ and the
strength of selection $|s|$.  Even though sites that violate the
infinite sites approximation may be very rare, they have a
disproportionate weight in estimates of these parameters and thus
can lead to errors of an order of magnitude or more (see below).
The method also erroneously infers positive selection in many
situations where selection is actually negative. Both of these
problems arise across a broad and biologically relevant range of
parameters (Fig. 1).

In addition to the systematic biases in the estimates it produces,
the PRF method also wastes data. In particular, it makes no use of
the information contained in the monomorphic sites of a sample.
Given a $\gamma$ and $\thetaseq$, the PRF method predicts the
number of sites $y_i$ that will exhibit $i$-fold polymorphisms, for
$i$ between $1$ and $n$, the sample size.  The PRF method makes no
prediction about the number of monomorphic sites (more precisely
the infinite sites approximation assumes that there is an infinite
set of monomorphic sites among which the polymorphism is found).
Thus if we had two sets of sequences with the same configuration of
polymorphic sites but different numbers of monomorphic sites, ML
estimation based on the PRF method would infer identical
(per-sequence) mutation rates and selective coefficients. This
holds despite the fact that monomorphic sites certainly contain
information: a larger number of monomorphic sites reflects a lower
mutation rate or stronger selection or both.

Two other assumptions of the PRF model are worth mentioning.  First,
the PRF method assumes that all sites experience the same selective
pressure.  In practice, this means that the PRF estimate of the
selection strength $s$ is actually some sort of weighted average of
the selective forces acting on the sites analyzed.  For example,
since any site with $|s| \gg \frac{1}{N}$ will almost always be
monomorphic, and data on monomorphism is ignored, such sites will
not influence the PRF ML estimate of the ``average" $s$.  But beyond
this, the weighting in this average is unclear, and hence it is
unclear what the PRF ML estimate of $s$ really represents.  We
explore this issue of variable selective pressures in more detail
below.

Finally, the PRF framework assumes that, once a mutation fixes at a
particular site, additional mutations at the site will experience the
same selective coefficient as the original mutation. For example, if a
mutation from nucleotide $A$ to $C$ fixes despite a selective
disadvantage $s$, new mutations at this site (e.g. from $C$ back to $A$)
are assumed to again have selective \textit{dis}advantage $s$. This
unrealistic assumption arises because the PRF focuses on a steady state
distribution of mutant lineages without references to the sites at which
they occur --- the fixation of old mutants is simply assumed to balance
the continuous generation of new ones, but not to change the selective
advantage of future mutations.  There is no reference to the
evolutionary dynamics at individual sites.  Although unrealistic for
negative selection, this approach has some advantages.  In particular,
it makes modeling positive selection straightforward: positive selection
is simply a constant flux of new mutations that increase the fitness,
and a steady state distribution of their frequencies is well defined.

\section{A Per-Site Poisson Random Field Model of Polymorphisms}

The problems with the PRF method described above can all be resolved by
replacing it with an explicit model of evolution at each site. We
develop such a model in the following section. First, however, we
describe in this section a method that retains the basic PRF framework,
but corrects some of the problems associated with the infinite-sites
assumption, and takes full advantage of the information provided by the
frequencies of all possible configurations at a site.

The basic idea behind this modified approach is to recast the PRF
framework on a \emph{per-site} basis.  We describe the steady state
frequency distribution of mutant lineages at a given site.  From this,
we can calculate the probability that a sample of $n$ individuals will
contain any configuration of mutants at that site.  As in the original
PRF method, we retain the assumption of free recombination, so that the
DNA sequence is a collection of independent sites. Thus our per-site
analysis leads directly to ML estimation of mutation rate and selection
strength.

We begin by recasting the PRF expression for the steady state
distribution of mutant lineages to describe the frequencies of
mutant lineages at a single site.  At a given site, we have \eon
f(x) = \thetasite \frac{1 - e^{-2 \gamma (1-x)}}{1 - e^{-2 \gamma}}
\frac{1}{x(1-x)}, \eoff where $\thetasite$ is the \emph{per-site}
value, $\thetasite = 2 N \mu$.  Using this formula to describe
multiple lineages at a single site is somewhat peculiar, because
this result assumes that all mutant lineages behave independently
of one another. Clearly this is not strictly true, since the mutant
lineages are segregating at the same site.  However, provided two
mutant lineages rarely achieve simultaneous high frequencies in the
population, then the assumption of independent mutant lineages is a
good approximation.  This assumption of non-interacting mutant
linages will often  hold  even when the other aspects of the
infinite-sites approximation are violated.

Analogous to the original PRF method, at \emph{a single site} the number
of mutant lineages which are observed $i$ times in a sample of $n$
sequences (``$i$-fold mutant lineages'') is Poisson distributed with
mean \eon F(i) = \thetasite \int_0^1 \frac{1-e^{-2
\gamma(1-x)}}{1-e^{-2 \gamma}} \frac{1}{x(1-x)} {n \choose i} x^i
(1-x)^{n-i} dx. \eoff  Based on this, we can calculate the
probability of any particular polymorphism configuration at a
site.

We begin by describing this calculation in the unfolded case. The
probability that a site is monomorphic is just the probability that
no $i$-fold mutant lineages are found at that site, for all $i$
between $1$ and $n-1$. This is \eon \pr{mono} = \pr{(n,0,0,0)} =
e^{-F(1)} e^{-F(2)} \ldots e^{-F(n-1)} \equiv M.  \eoff The
probability that a site exhibits a $(n-1,1,0,0)$ configuration is
the probability that a single $1$-fold mutant lineages is sampled,
and no $2$-fold or higher lineages are found, \eon \pr{(n-1,1,0,0)}
= F(1) M. \eoff  The probability of exhibiting an $(n-2,2,0,0)$
configuration is more complex. This configuration could arise from
a single $2$-fold sampled lineage (as assumed under the standard
PRF method) \textit{or} it could arise from two $1$-fold sampled
mutant lineages which happen to involve mutations to the same
nucleotide.  Hence its probability is \eon \pr{(n-2,2,0,0)} =
\frac{F(1)^2}{2!} M \frac{1}{3} + F(2) M, \eoff where the factor of
$\frac{1}{3}$ is the probability that two mutations result in  the
same nucleotide.  This expression assumes that mutations between
all possible nucleotides are equally likely --- the obvious
generalization applies when there are mutational biases, which we
do not discuss further. Similarly the probability of an
$(n-2,1,1,0)$ configuration is \eon \pr{(n-2,1,1,0)} =
\frac{F(1)^2}{2!} M \frac{2}{3}. \eoff The probabilities of more
complex configurations can be calculated in a similar way.  The
probability of an $(n-4,4,0,0)$ configuration, for example, is the
probability of four $1$-fold lineages to the same nucleotide plus
the probability of two $1$-fold lineages and a $2$-fold lineages,
plus the probability of two $2$-fold lineages, plus the probability
of one $1$-fold lineage and one $3$-fold lineage, plus the
probability of a single $4$-fold lineage.  We have \eon
\pr{(n-4,4,0,0)} = M \left[ \frac{F(1)^4}{4!} \frac{1}{3^3} +
\frac{F(1)^2 F(2)}{2!} \frac{1}{3^2} + \frac{F(2)^2}{2!}
\frac{1}{3} + F(1)F(3) \frac{1}{3} + F(4) \right]. \eoff  In
general, the probability of a particular configuration is given by
the sum of the probabilities of all possible partitions of $n$ that
lead to that configuration.

In the folded case, these calculations become even more complex.
The probability of a $(12,2,0,0)$ folded configuration, for
example, includes the probability of a single $12$-fold sampled
lineage, as well as two $6$-fold sampled lineages, and so on.
Thousands of terms may arise in the expression for the probability
of a particular configuration, even for moderate values of $n$.  We
do not quote any of these results here, but rather we have
developed a computer program to output symbolic expressions for the
probabilities of all possible folded as well as unfolded
configurations, for a given sample size $n$ (available on request).

These probabilities of site configurations form the basis of
maximum likelihood parameter estimation.  The probability of a data
set with $L$ total sites, including $L_{a,b,c,d}$ sites in
configuration $(a,b,c,d)$ is given by \eon \frac{L!}{\prod
L_{a,b,c,d}!} \prod \pr{a,b,c,d}^{L_{a,b,c,d}}, \eoff where the
products are over all possible configurations. Given a particular
data set, we maximize this probability over $\thetasite$ and
$\gamma$ to find the ML estimate of these parameters.  In the
original PRF method, this ML estimation is particularly simple,
because the ML estimate for $\theta$ can be expressed analytically
in terms of the ML estimate for $\gamma$, leaving a one-dimensional
numerical maximization procedure to estimate $\gamma$. In our
per-site PRF method, however, a full two-dimensional numerical
maximization is required to find the ML estimates of $\thetasite$
and $\gamma$.

Our per-site PRF method relaxes most of the consequences of the
infinite-sites approximation inherent in the original PRF estimation
procedure.  We allow for the possibility that multiple mutant
lineages contribute to the polymorphism observed at a single site.
Thus we avoid the systematic underestimation of $\theta$ and
$\gamma$, and incorrect inference of positive selection, that affect
the traditional PRF (see Numerical Simulations).  This new method
also uses all of the data available in the sample, including the
number of monomorphic sites and the differences between
$(n-2,2,0,0)$ and $(n-2,1,1,0)$ sites.   It does still retain one
aspect of the infinite-sites approximation: it assumes that mutant
lineages segregating at the same site are independent (i.e. they do
not interfere with each other).  This is never strictly true, but is
a good approximation unless multiple mutant lineages reach high
frequency at a given site at the same time. Note that because of
this assumption, the probabilities of all possible configurations
$(a,b,c,d)$ described above do not precisely sum to unity, because
our approach allows a typically small but nonzero probability of
multiple mutant processes adding to more than $n$ sampled
individuals.  Since the no-interference approximation is always
valid when the infinite-sites approximation is, and it holds in many
situations where the infinite-sites approximation fails, our revised
sampling method extends the applicability of the PRF framework and
fixes many of its problems.

\section{A Per-Site Diffusion Model of Polymorphisms}

In this section, we describe a method that shifts fundamentally from
the PRF framework.  Rather than studying the distribution of the
frequencies of mutant lineages, we focus on the evolutionary
dynamics at each individual site, without keeping track of
individual mutant lineages. We develop this into a
maximum-likelihood estimation of $\gamma$ and $\theta$ from
polymorphism data, which requires neither the infinite-sites or
no-interference approximation described above. As in the original PRF
method, we assume free recombination between sites.

At an individual site, we imagine that one nucleotide is preferred,
and the other three have the same fitness disadvantage $s$ ($s <
0$).  We assume that mutations occur at rate $\mu$, and hence at
rate $\mu/3$ between any two specific nucleotides (i.e. no
mutational biases).  These assumptions simplify the discussion, but
are not essential.  In fact, one advantage of this approach is that
these assumptions can be easily relaxed with obvious generalizations
(noted below).

We can analyze the process of mutation, selection, and drift at a
single site with a three-dimensional diffusion approximation, and
calculate the joint steady-state probability distribution of the
frequencies of the four possible nucleotides at the site.  This
then leads naturally to the likelihood of any configuration of
polymorphism data at the site as a function of $\gamma$ and
$\thetasite$, and hence to a ML estimation of these parameters from
data.  Alternatively, we can sacrifice some of the information in
the data, and reduce the computational complexity of the problem by
treating all three disfavored nucleotides as a single class.  Such
a treatment reduces to a standard one-dimensional diffusion process
whose steady state probability distribution describes the frequency
of the preferred nucleotide versus the sum of the frequencies of
the disfavored ones; this treatment is essentially a steady-state
version of \citet{williamson05}. This approach discards some of the
information in the data (e.g. not making use of the difference
between $(12,1,1,0)$ and $(12,2,0,0)$ sites), but it is
computationally simpler.  We begin by describing the
one-dimensional method, and then turn to the three-dimensional
method.

\subsection{The One-Dimensional Diffusion Model}

We begin by describing a simplified diffusion approach that calculates
the frequency distribution of favored versus disfavored nucleotides.  As
noted above, we will for simplicity assume that one nucleotide is
preferred, and the other three nucleotides are disfavored.  We denote
the sum of the frequencies of the three \emph{disfavored} alleles by
$x$; the frequency of the preferred nucleotide is $1-x$.

We assume that mutation, selection, and random drift occur at each site
according to standard Wright-Fisher dynamics.  Thus the probability
distribution of $x$ can be described by the diffusion equation \eon
\pd{}{t} f(x,t) = \frac{1}{2} \frac{\partial^2}{\partial x^2} \left[
v(x) f(x,t) \right] + \pd{}{x} \left[ m(x) f(x,t) \right], \eoff where
$f(x,t)$ is the probability that the disfavored nucleotides sum up to
frequency $x$ at time $t$, $s$ is the selection coefficient against the
disfavored nucleotides ($s < 0$), $\mu$ is the per-site mutation
rate per individual per generation, and \begin{eqnarray} m(x) & = & s x
(1-x) + \mu (1-x) - \frac{\mu}{3} x \\ v(x) & = & \frac{x(1-x)}{N}.
\end{eqnarray}  This diffusion equation is well-known and has the
steady-state solution \eon f(x) = C x^{\thetasite-1}
(1-x)^{\thetasite/3-1} e^{2\gamma x}, \eoff where $C$ is a ($\thetasite$
and $\gamma$-dependent) normalization factor, and as before $\thetasite
\equiv 2 N \mu$ and $\gamma \equiv N s$.

If the frequency of disfavored nucleotides at a site equals $x$, the
probability that we find $i$ such nucleotides in a sample of $n$
individuals is ${n \choose i} x^i (1-x)^{n-i}$.  Averaging over $x$, the
overall probability that we sample $i$ disfavored nucleotides at a given
site is \eon F(i) = {n \choose i} \int_0^1 C x^{\thetasite+i-1}
(1-x)^{\thetasite/3 +n-i-1} e^{2 \gamma x}\ \text{d}x. \eoff This integral,
including the calculation of the normalization factor $C$, can be solved
analytically.  We find \eon F(i) = {n \choose i} \frac{\Gamma
(n-i+\thetasite/3) \Gamma (i + \thetasite) {}_1 F_1 (i + \thetasite, n +
4 \thetasite/3, 2 \gamma)}{\Gamma (\thetasite/3) \Gamma (\thetasite)
{}_1 F_1 (\thetasite, 4 \thetasite/3, 2 \gamma)}, \eoff where $\Gamma$
is Euler's Gamma function and ${}_1 F_1$ is a hypergeometric function.

The expression above leads immediately to a maximum likelihood method for
estimating $\gamma$ and $\thetasite$ in the unfolded case -- i.e. when
the identity of the preferred nucleotide at each site is known.  In a
sample of $n$ sequences each of length $L$, we count the number of
sites at which $i$ disfavored nucleotides are sampled, $y_i$, for $0\leq
i \leq n$.  Since all sites are assumed independent, each with the
polymorphism frequency distribution described above, the
likelihood of the data given the parameters is \eon y_u(\thetasite, \gamma)
= \frac{L!}{\prod_{i=0}^{n} y_i!} \prod_{i=0}^n F(i)^{y_i}.  \eoff For
any set of polymorphism data, it is straightforward to  maximize this
function numerically, producing ML estimates of $\thetasite$ and
$\gamma$.  As with the per-site PRF method, this procedure involves a
two-dimensional maximization routine.

When we do not know which nucleotide is preferred at each site, we
must use the ``folded'' version of the data.  This presents
difficulties. Imagine a site with an $(a,b,c,d)$ polymorphism
configuration.  We might naively suppose that since any of the four
nucleotides could be the preferred one, the probability of this
data is simply $F(b+c+d)+F(a+c+d)+F(a+b+d)+F(a+b+c)$.  However,
this is not the case. For example, if $a$ is indeed the preferred
nucleotide, then $F(b+c+d)$ does not equal the probability that the
three disfavored nucleotides will form a $(b,c,d)$ configuration.
Rather, it equals the sum of the probabilities that the three
disfavored nucleotides will form a configuration $(i,j,k)$, summed
over all $i,j,k$ triplets that sum to $b+c+d$.  This is a serious
problem, because the difference between $F(b+c+d)$ and the
probability of the data depends on $b+c+d$, and hence it is not
identical for all four possible preferred nucleotides.  Simply
assuming that the most common nucleotide is the preferred one
\cite{hartlsawyer94} is a reasonable approach to folded fits.  But
this approach will be inaccurate for sites where the most common
nucleotide is not overwhelmingly so; when this situation describes
a substantial fraction of sites, the method will fail.  As a
result, the one-dimensional diffusion framework does not allow for
a rigorous ML estimate of parameters with folded data. The exact
same problem also arises in the traditional PRF method, although it
tends to be obscured by the other problems with that method.

In order to perform rigorous ML fits to folded frequency data we
must turn to a three-dimensional diffusion method, which we will
now discuss.

\subsection{The Three-Dimensional Diffusion Model}

Rather than considering all disfavored nucleotides as a single class, we
can instead keep track of the evolutionary dynamics of all four possible
nucleotides at a site.  In order to do so, we assume the standard
four-allele Wright-Fisher dynamics, with mutation at rate
$\frac{\mu}{3}$ between any two particular nucleotides, and selection
acting with strength $s$ against the three disfavored nucleotides.  The
dynamics can then be described by a three-dimensional diffusion equation
for the joint distribution of the frequencies of the three disfavored
alleles $x_1, x_2$, and $x_3$, $f(x_1, x_2, x_3, t)$ (where the
preferred allele has frequency $x_0 = 1-x_1-x_2-x_3$).  We have \eon
\pd{}{t} f(x_1,x_2,x_3,t)  = \frac{1}{2} \sum_{i=1}^3 \sum_{j=1}^3
\frac{\partial^2}{\partial x_i \partial x_j} \left[ V_{ij} \cdot f
\right] - \sum_{i=1}^3 \pd{}{x_i} \left[ M_i \cdot f \right], \eoff
where \begin{eqnarray} M_i & = & (1+s) x_i \left(1-\sum_{j=1}^3
x_j\right) + \frac{\mu(1-4x_i)}{3}\\ V_{ii} & = & \frac{x_i(1-x_i)}{N}\\
V_{ij} & = & -\frac{x_i x_j}{N} \ \ \ \ \ \ \ \ \ (i\ne j).
\end{eqnarray} This is a somewhat less well-known
diffusion equation \cite{watterson77, wright49}; the steady-state
solution is \eon f(x_1,x_2,x_3) = C \left[ x_1 x_2 x_3 (1 - x_1 -
x_2 - x_3) \right]^{\L} e^{2 \gamma \left( x_1 + x_2 + x_3
\right)}, \eoff where $C$ is a normalization factor, and we have
defined $\L = \thetasite/3-1$.

Given the frequencies $x_0$, $x_1$, $x_2$, and $x_3$ of nucleotides
in the population, the probability of sampling a site in an
\emph{unordered} configuration $(n_0,n_1,n_2,n_3)$ in a sample of
$n$ individuals (adopting the convention that the first nucleotide
listed is the preferred one) is just the multinomial probability
\eon \frac{n!}{n_0! n_1! n_2! n_3!} (1-x_1-x_2-x_3)^{n_0} x_1^{n_1}
x_2^{n_2} x_3^{n_3}. \eoff Averaging over $f$, we therefore find
that the probability of sampling a site in an unordered
$(n_0,n_1,n_2,n_3)$ configuration is
\begin{eqnarray} \pr{n_0,n_1,n_2,n_3} & = & C \frac{n!}{n_0! n_1!
n_2! n_3!} \int_0^1 \int_0^{1-x_1} \int_0^{1-x_1-x_2} \times \\ & &
e^{2 \gamma \left( x_1 + x_2 + x_3 \right)} x_1^{\L+n_1}
x_2^{\L+n_2} x_3^{\L+n_3} \left[1-x_1-x_2-x_3 \right]^{\L+n_0} dx_3
dx_2 dx_1 . \label{unordered}
\end{eqnarray}

In some applications, we will know which of the nucleotides is
preferred at each site --- i.e. the unfolded case. In this case,
the probability of finding a site in an \emph{ordered} unfolded
configuration $(a,b,c,d)$ (where by convention $a$ is the number of
individuals which have the preferred nucleotide and $b \geq c \geq
d$), is \eon P^u_{a,b,c,d} = \sum_{\{ n_0, n_1, n_2, n_3 \}}
\pr{n_0,n_1,n_2,n_3}, \eoff where the sum is over all unordered
configurations $(n_0,n_1,n_2,n_3)$ that give rise to the ordered
unfolded configuration $(a,b,c,d)$.

In other cases, we do not know which of the nucleotides is preferred
at each site -- i.e. the folded case.  Here, the probability of
sampling a site in the ordered configuration $(a,b,c,d)$ is just the
sum of the probabilities assuming that each of the four possible
nucleotides is preferred.  As before, we adopt the convention that
ordered folded configurations are written as $(a,b,c,d)$ with $a
\geq b \geq c \geq d$.  The probability of a folded configuration
$P^f_{a,b,c,d}$ is then \eon P^f_{a,b,c,d} = \sum_{\{ n_0, n_1, n_2,
n_3 \}} \pr{n_0,n_1,n_2,n_3}, \eoff where in this case the sum is
over all unordered configurations that give rise to the ordered
folded configuration $(a,b,c,d)$.

For either folded or unfolded data, given $n$ samples of a sequence
$L$ sites long, with $L_{a,b,c,d}$ sites in an $(a,b,c,d)$
polymorphism configuration, the likelihood of the data is \eon
\like (\thetasite,\gamma) = \frac{L!}{\prod L_{a,b,c,d}!} \prod
\left[P_{a,b,c,d}\right]^{L_{a,b,c,d}}, \eoff where the products
are taken over all possible configurations $(a,b,c,d)$ and
$P_{a,b,c,d}$ is the folded or unfolded probability defined above.
We can numerically maximize this function to find the ML estimates
of $\thetasite$ and $\gamma$.

In practice, the ML estimation procedure described above is
difficult to implement, because of the triple integral in the
definition of $\pr{n_0,n_1,n_2,n_3}$ (as well as that implicit in
the definition of the normalization constant $C$). This integral
cannot be solved exactly, and it is difficult to evaluate
numerically  because the integrand may diverge (though the integral
itself converges) near the boundary of the simplex over which it is
integrated.  We adopt a hybrid method to simplify the evaluation of
this triple integral.  Near the boundary of the simplex, we Taylor
expand the integrand and integrate it analytically.  Away from the
boundary, the integrand is well-behaved and standard numerical
integration has no difficulties.  This approach is most easily
achieved by making the substitutions $y = x_3/(1-x_1-x_2)$ and $z =
x_2/(1-x_1)$.  On doing so, we can rewrite the integral as
\begin{eqnarray} \pr{n_0,n_1,n_2,n_3} & = & C \frac{n!}{n_0! n_1!
n_2! n_3!} \int_0^1 \int_0^1 \int_0^1 \times \\ & & \exp \left[ 2
\gamma \left(x + y + z - x y - z y - x z +xyz\right) \right]
x^{n_1+\L} z^{n_2+\L} y^{n_3+\L} \times \nonumber \\ & & \left( 1-x
\right)^{n_0 + n_2 + n_3+ 3\L + 2} \left( 1-z \right)^{n_0 + n_3 +
2\L + 1} \left( 1-y \right)^{n_0 + \L} \nonumber.
\end{eqnarray}
This expression is much easier to handle than our original expression,
because the three integrals can be done in arbitrary order.  We divide
each of the three integrals into three pieces: one from $0$ to $\delta$,
one from $\delta$ to $1-\delta$, and one from $1-\delta$ to $1$.  Thus
the triple integral is split into $27$ total terms.  For each of the
integrals from $0$ to $\delta$ or $\delta$ to $1-\delta$, we Taylor
expand the integrand in the integration variable, and we solve the
integral analytically.  All of the remaining integrals, from $\delta$ to
$1-\delta$, are done numerically.  We must choose $\delta$ large enough
that we can perform the numerical integrals quickly, but not so large
that the Taylor expansions used for the analytical parts become invalid.
For these Taylor expansions, we need $\delta \ll \frac{1}{2 \gamma}$,
$\delta \ll 1$, $\delta \ll \frac{1}{3\L+2}$, $\delta \ll \frac{1}{2 \L
+ 1}$, and $\delta \ll \frac{1}{\L}$.  For the computational analysis
described in this paper, we choose whichever of these conditions is most
restrictive and set $\delta$ to be one-tenth of the most restrictive
requirement.  We find that this choice of $\delta$ is sufficiently small
to provide accuracy in the analytical parts of the integrals, but large
enough to enable quick numerical integration on the interior of the
simplex.

\subsection{Comparison between the PRF and diffusion methods}

Both our one- and three-dimensional diffusion approaches relax
all of the infinite-sites assumptions of the PRF framework.  This
includes the interference between mutant lineages segregating at the
same site, which even the per-site PRF method mishandles.  Thus, the
diffusion approach contains none of the biases associated with
infinite-sites approximation that plague the traditional PRF and, to a
lesser degree, the per-site PRF.  The diffusion approach also provides a
clear and concrete model of the evolution at each site.  This contrasts
with the PRF method, which makes the unrealistic assumption that if a
deleterious allele fixes at a site, further mutations are again
deleterious.

The diffusion method is also easily extendable to more complex
evolutionary situations.  For example, we can explore different
selective costs for different nucleotides, more than one preferred
nucleotide, or mutational biases.  These possibilities lead to obvious
modifications of the diffusion equations and their solutions, and hence
to the maximum likelihood estimation.  It is also straightforward to
investigate balancing selection or the effects of dominance:  these lead
to well-understood modifications to the diffusion equations and their
steady state solutions \cite{ewensbook}.  In the PRF framework, by
contrast, such generalizations are much more complex.  In particular,
balancing selection is impossible to analyze under the PRF framework,
because it leads to mutant lineages reaching stable intermediate
frequencies in the population. As a result, the generation of new
mutations is not balanced by the extinction or fixation of older ones,
and hence no steady state distribution of lineage frequencies exists.

However, the diffusion approach is not without drawbacks.  The
one-dimensional version does not make use of all of the data available
in the observed polymorphism spectrum.  The three-dimensional version
does use all the data, but its implementation requires tedious numerical
integration routines.

The diffusion method also cannot naturally handle positive
selection. The steady state evolutionary dynamics at a site are
always dominated by the preferred nucleotide (or nucleotides), with
negative selection acting against polymorphisms for the disfavored
nucleotides.  At the level of an individual site, positive selection
is a process that is intrinsically out of steady state:  the spread
of a favorable nucleotide before it becomes fixed.  The PRF method
handles this by positing that there are a wide array of sites which
have the potential for positive selection, and assuming that a
steady state \emph{across sites} of these selective sweeps is
maintained.  It should be possible to address positive selection
within the diffusion framework by changing the boundary conditions
of our diffusion equations, so that the fixation of one mutant
lineage shifts the selective landscape so that further mutations are
now favored.  Formally, any probability flowing into $x=1$ is
``absorbed'' and moved to $x=0$.  Alternatively, one could use full
time-dependent solutions to the diffusion equations to study
positive selection.  However, we do not pursue these approaches in
this paper, and instead focus our study on the case of negative
selection.

\section{Variable selection pressures across sites}

Both the original PRF method as well as the three per-site methods
we have proposed in this paper assume that all sites experience the
same selective pressure.  In reality, we expect that there is some
distribution of selective pressures across sites.
\citet{hartlsawyer94} suggest that in this case the ML estimate of
$\gamma$ from their PRF method reflects an ``average'' selection
pressure across the sites. In fact, the ML $\gamma$ reflects a
\emph{weighted} average of the actual $\gamma$'s across the sites,
but the nature of this weighting is not understood. Almost no
weight is given to sites at which $|\gamma| \gg 1$, because these
sites will almost always be monomorphic and hence ignored by the
original PRF method. It is unclear how sites with different values
of $\gamma$ of order $1$ will be weighted, or how the presence of
some effectively neutral sites ($|\gamma| \ll 1$) will change the
ML estimate.

The issue of variable $\gamma$ across sites is of even greater concern
for the methods we have proposed, because these methods make use of the
monomorphism data.  If some number $L_l$ sites are effectively lethal
(i.e. $|\gamma| \gg 1$), these sites will increase the number of
monomorphic sites, which will tend to depress our ML estimate of
$\thetasite$ and increase our estimate of $|\gamma|$.  Fortunately, our
methods are able to use the monomorphism data in order to investigate
the number of lethal sites, $L_l$, or more generally to infer a full
distribution of selection pressures across sites.

Since all of the methods we have proposed are defined at a per-site
level, it is straightforward to assume that there are multiple
different classes of sites with different values of $\gamma$.  We
can posit that there are $k$ classes of sites.  Each class is
represented by $L_k$ of $L$ total sites, and has its own value of
$\gamma$, which we call $\gamma_k$.  The probability that a site is
in an $(a,b,c,d)$ configuration is then \eon \pr{a,b,c,d} =
\sum_{j=1}^k \frac{L_j}{L} \pr{a,b,c,d}^j (\gamma_j, \thetasite),
\label{new30} \eoff where $\pr{a,b,c,d}^j (\gamma_j, \thetasite)$
is the probability that site with parameters $\gamma_j$ and
$\thetasite$ is in the configuration $(a,b,c,d)$.  This expression
is correct both for the per-site PRF and per-site diffusion
approaches --- so we can apply this method regardless of which
approach we are using.

Given our new definition of $\pr{a,b,c,d}$, we can construct the
folded or unfolded likelihood of the overall polymorphism data set in
exactly the same way as before.  This likelihood function now depends on
$2k+1$ parameters: $\thetasite$, the $\gamma_j$, and the $L_j$.  We can
find ML estimates of all of these parameters using a multidimensional
numerical maximization of the likelihood function.  By choosing $k$, we
determine the resolution at which we measure the distribution of values
of $\gamma$ across sites. Naturally, the larger the $k$ we choose, and
hence the greater the resolution on $\gamma$, the more data we require
to obtain accurate estimates of the individual $L_j$ and $\gamma_j$.

Rather than estimating both the $L_j$ and the $\gamma_j$, we could
instead posit that there are several classes of mutations with
different \emph{pre-specified} $\gamma_j$, and estimate only the
values of $L_j$. In other words, we ask what fraction of sites have
different values of selective constraints.  We describe here one
particularly important example of a hybrid between these two
procedures, with two classes of sites ($k=2$). Rather than fitting
an ML estimate of $\gamma$ to both classes, we assume that one class
of sites is unable to evolve: mutations at these sites are lethal
(more precisely, they have $|\gamma| \gg 1$). We wish to calculate
the number of lethal sites, and the average selective pressure on
the remaining, non-lethal sites.  Thus we have three parameters: the
mutation rate $\thetasite$, the number of lethal sites $L_2$, and
the strength of selection $\gamma$ acting at the other $L_1 = L-L_2$
sites. The probability that a site is monomorphic is given by \eon
\pr{mono} = \frac{L_2}{L} + \frac{L_1}{L} \pr{mono}^1. \label{new31}
\eoff Here $\pr{mono}^1$ is the probability that a site with
strength of selection $\gamma$ and mutation rate $\thetasite$ will
be monomorphic, as defined by either the per-site PRF or per-site
diffusion approach (whichever method we are using). The probability
that a site is in a non-monomorphic $(a,b,c,d)$ configuration is
\eon \pr{a,b,c,d} = \frac{L_1}{L} \pr{a,b,c,d}^1. \label{new32}
\eoff Now we can write the likelihood of the data in the usual way.
This results in a three-dimensional ML problem. However, we can
simplify the problem by first maximizing $L_2$ given $\gamma$ and
$\theta$. We find that the ML estimate of $L_2$ is \eon \hat L_2 =
\frac{L_{mono} - L \pr{mono}^1}{1-\pr{mono}^1}, \label{MLL2} \eoff
where $L_{mono}$ is the number of monomorphic sites in the data and
$\pr{mono}^1$ is the probability a non-lethal (i.e. an $L_1$) site
is monomorphic. Substituting this value for $L_2$, we are left with
a two-dimensional maximization problem in $\gamma$ and $\thetasite$,
similar to the original situation.

It is worth exploring how this procedure for estimating the number
of lethal sites utilizes the data. As we now show, this procedure
is equivalent to ignoring the monomorphism data when finding the
maximum-likelihood estimates of $\gamma$ and $\thetasite$ for the
non-lethal sites. The likelihood of the data ignoring monomorphic
sites is \eon L(\thetasite, \gamma) = \frac{L_p!}{\prod
L_{a,b,c,d}} \prod \left[ \frac{\pr{a,b,c,d}^1}{1-\pr{mono}^1}
\right]^{L_{a,b,c,d}}, \label{ignoremono} \eoff where $L_p$ is the
total number of non-monomorphic sites and the products are over all
configurations of non-monomorphic sites. After finding ML estimates
of $\gamma$ and $\theta$ from this monomorphism-ignoring likelihood
function, the procedure then calculates the number of monomorphic
sites that would be expected given $\gamma$ and $\thetasite$.  This
is $L_1 \pr{mono}^1 = (L - L_2) \pr{mono}^1$. We then estimate the
number of lethal sites $L_2$ as the difference between the observed
number of monomorphic sites $L_{mono}$ and the number that would be
predicted if all sites were of the $L_1$ variety, $\hat L_2 =
L_{mono} - (L - \hat L_2) \pr{mono}^1$. Rearranging this
expression, we see that it is identical to \eq{MLL2} above.  And
indeed, plugging \eq{new31}, \eq{new32}, and \eq{MLL2} into
\eq{new30} yields \eq{ignoremono}.

Thus, this procedure ignores the monomorphism data when calculating
$\gamma$ and $\thetasite$ (at the non-lethal sites), and it instead
uses the monomorphism data to infer one aspect of the distribution
of $\gamma$ across sites --- specifically, the number of lethal
sites. Since the original PRF method also ignores monomorphism data,
we obtain this information on the distribution of $\gamma$ for
``free,'' relative to the power of the original method, simply by
shifting to the per-site model.  If desired, we can also posit that
there are a number of sites $L_3$ which are effectively neutral
(i.e. with $|\gamma| \ll 1$), and estimate $L_3$.  This would devote
some part of the data describing polymorphisms at intermediate
frequencies to estimating $L_3$, though the division in which data
are used for estimating which parameters is not as sharp as in the
case of lethal mutations.  From this procedure we could estimate the
number of effectively lethal sites, the number of effectively
neutral ones, and the ``weighted average'' selection pressure acting
on the remaining sites. If more resolution is desired, and enough
data are available, we can increase the number of classes of sites
and obtain ML estimates of the numbers of sites in each class and
the selection pressure acting on each class.

\section{Accuracy of inference techniques}

Using data generated from Wright-Fisher simulations, we have tested the
inferential accuracy of the PRF method as well as the accuracy of our
three alternative methods. The Wright-Fisher model (or, more precisely,
its diffusion limit) forms the basis of the PRF method, and it is
therefore the appropriate simulation framework for testing the method.

All simulations assumed a constant population of $N=1000$ haploid
individuals. Each of $L=1000$ sites, simulated independently, could
assume one of four states: a, c, t, or g. One state is assigned
fitness 1, and the other three states fitness $1+s$. Mutations
occurred at rate $\mu$ per site. The allele frequencies evolved
according to the standard Wright-Fisher Markov chain
\cite{ewensbook}.  Each simulation was run for at least $10/\mu$
generations, so as to ensure relaxation to steady-state. At the end
of the simulation, $n=14$ individuals were sampled from the
population and the polymorphism frequency spectrum was recorded.
For `unfolded' fits, the identity of the preferred nucleotide was
retained, whereas this information was discarded for `folded' fits.
We chose to consider samples of size $n=14$ in order to facilitate
comparison with \citet{hartlsawyer94}.

We performed simulations over a wide range of parameter values. We
considered five different values of $\thetasite$: 0.05, 0.1, 0.5,
1.0, and 5.0. For each value of $\thetasite$, we performed one
simulation at each of 17 different values of $\gamma$, ranging from
$\gamma=-10.0$ to $\gamma=-0.1$. For each set of simulation
parameters $(\gamma,\thetasite)$, once the simulated polymorphism
data had been generated, ML parameter estimates
$(\hat{\gamma},\hat{\thetasite})$ were obtained by numerical
maximization of the likelihood function, as specified by the
original PRF model, the per-site PRF model, the one-dimensional
diffusion model, or the three-dimensional diffusion model. A 95\%
confidence interval for $\gamma$ was constructed according to
\citet{bustamante01}: the interval includes those values of
$\gamma$ within $0.5 \chi^2_{1,0.95}$ likelihood units from
$\hat{\gamma}$. The estimated parameters shown in Figures 1-5 are
somewhat `jagged,' because the inference methods have been applied
to a single draw of $n=14$ sequences for each set of simulation
parameters, as opposed to averaging over many such draws.

As discussed above, the original PRF model disallows multiple mutant
lineages at a site. Therefore, when a site sampled from the simulated
data exhibited more than two types of segregating nucleotides, the
frequencies of all unpreferred nucleotides were summed to represent the
frequency of the `mutant' type, as suggested by \citet{hartlsawyer94}.
In addition, when fitting folded data using the PRF method, the most
common nucleotide was assumed to be the ancestral type, as suggested by
\citet{hartlsawyer94}. This approach is not entirely accurate, as
discussed above, but it is probably the best option available within the
original PRF framework.

Fig. 2 compares the accuracy of estimated selection pressures using the
original PRF method versus the the one-dimensional diffusion method.
Fig. 3 shows the same type of comparisons over a range of mutation
rates, including also the modified, per-site PRF method. The original
PRF method systematically underestimates the strength of negative
selection, by as much as a factor of 10. In fact, the PRF method
strongly rejects the true parameters in over 85\% of the cases. In
addition, the PRF method often erroneously infers strong positive
selection when in fact mutants are under negative selection.  These
problems are more severe when the mutation rate is large, but also occur
for small mutation rates. The smallest mutation rate shown in Fig. 3 is
one-half the mutation rate estimated for bacterial genes
\cite{hartlsawyer94}.  The per-site version of the PRF method that we
have developed corrects the most severe problems of the standard PRF
method, but it too exhibits systematic biases, especially when selection
is weak and the mutation rate large (Fig. 3). The one-dimensional
diffusion method that we have developed provides accurate and unbiased
estimates of $\gamma$ over the full range of selective pressures and
mutation rates (Fig. 3). Like its one-dimensional counterpart, the
three-dimensional diffusion method also provides accurate and unbiased
estimates over the full range of simulated parameters (not shown).

For folded data, Fig. 4 shows the accuracy of inferred selection
pressures using the original PRF method, the per-site PRF method,
and the three-dimensional diffusion method. Again, the original PRF
method systematically underestimates the strength of selection, and
it also erroneously infers positive selection. As before, the
per-site PRF method corrects the most severe problems of the
original PRF method, but it exhibits systematic biases at large
mutation rates (Fig.  4). The three-dimensional diffusion method
that we have developed provides unbiased estimates of selection
pressures.  When selection is weak (\textit{i.e.} $|\gamma|<1$),
however, the confidence intervals on diffusion-based estimates of
$\gamma$ are appreciably larger in the folded case, compared to
the unfolded case (Fig. 3c versus Fig. 4c). This behavior makes
perfect sense: when selection is nearly neutral and the ancestor
state is unknown, the frequency distribution does not exhibit
sufficient skew to deduce the preferred nucleotide. As a result,
the diffusion-based estimator cannot distinguish between weak
positive and weak negative selection in the absence of information
on the preferred nucleotide (Fig. 4c). Thus, the confidence
intervals obtained under the folded diffusion technique properly
reflect our inability to estimate the selection pressure precisely
when selection is weak.

As shown in Figures 3 and 4, when selection is weakly negative the
original PRF method erroneously infers positive selection, regardless of
the mutation rate.  In fact, this problem occurs in the exact parameter
regimes that have been estimated from biological data. For example, on
the basis of $n=14$ sampled sequences each $367$-sites long,
\citet{hartlsawyer94} estimated $\gamma=-1.34$ and $\thetasite=0.0915$
for silent sites in a bacterial gene. If we simulate $367$ Wright-Fisher
sites under these parameters and sample $n=14$ sequences, we find that
the most likely parameters fit using the original PRF method are
$(\hat{\gamma},\hat{\theta_s})=(+18.45,0.067)$. This exercise
demonstrates that the PRF method not only infers the wrong sign of
selection, but it is also an inconsistent estimator under biologically
realistic parameters.

%

Fig. 5 shows the accuracy of estimated mutation rates using the original
PRF method, the per-site PRF method, and the diffusion methods. The
original PRF method systematically underestimates the mutation rate, by
a factor as large as 30. In the unfolded case, the tendency to
underestimate the mutation rate is stronger when selection is weak.
Estimates obtained using the per-site PRF method that we have developed
partly correct these problems, but still exhibit biases at large
mutation rates. Our diffusion-based methods provide accurate and
unbiased estimates of the mutation rate, for both folded and unfolded
data, across the full range of mutation rates and selection pressures
(Fig. 5).

The methods we have developed in this paper also allow us estimate the
distribution of selection pressures across sites. In one simple case
discussed above, we have presented a procedure for estimating the number
of lethal sites and the selective pressure operating on the remaining,
non-lethal sites in a gene. This procedure involves estimating $\gamma$
and $\theta$ on the basis of polymorphic sites alone, and thereafter
estimating the proportion of observed monomorphic sites that are lethal.
In order to assess the power and accuracy of this approach, Table 1
shows the predicted number of monomorphic sites in each of our
simulations, compared to the number of monomorphic sites actually
observed. Across a large range of selective pressures and mutation
rates, this approach typically estimates the number of (non-lethal)
monomorphic sites within a few percent. As a result, for a gene of
length $L=2000$ sites, one-half of which are lethal, our procedure will
accurately predict the number of lethal sites within a few percent; and
it will accurately predict the selection pressure on the remaining,
non-lethal sites.

The simulations and fits presented in this section reflect our intuitive
understanding of the assumptions underlying the PRF model versus the
 methods we have developed.  For example, the PRF method
disallows multiple mutant lineages at a site and should therefore lead
to systematic underestimates of the mutation rate and selective
strength, even when mutations are segregating at low frequency. Our
simulations and fits verify this behavior.  The per-site PRF method
corrects the most problematic aspects of infinite-sites assumption, but
it still leads to biased inferences at large mutation rates because it
neglects interactions among high-frequency mutant lineages.  Finally,
the one- and three-dimensional diffusion models avoid the infinite-sites
approximation altogether, as reflected by the accuracy of estimates
obtained across the full range of parameters.

\section{Discussion}

The Poisson Random Field model \cite{sawyerhartl92} and the associated
likelihood procedure for estimating parameters \cite{hartlsawyer94} are
perfectly valid when the assumptions underlying the method are met ---
namely, infinite sites, free recombination, and constant selective
pressure across sites. We have shown, however, that the PRF method leads
to severely incorrect inferences in practice, because in reality genes
contain a finite number of sites.  We have developed three new methods
that relax or remove the infinite-sites assumption.  These new methods
not only fix the problems associated with the PRF method, but they also
extend the types of inferences that can be drawn from polymorphism data
to include inferences on the distribution of selective pressures across
sites.

It may seem surprising that the infinite-sites approximation can
lead to such drastic errors in the mutation rates and selective
strengths inferred by the PRF method.  After all, sites at which
multiple mutant lineages are sampled are presumably very rare.
Multiple polymorphisms (i.e. three or more segregating nucleotides)
occur in only a few percent of sites for much of the data analyzed
by the PRF method (e.g. \citet{hartlsawyer94}).  Yet despite their
rarity, these sites have a large impact on maximum likelihood
estimation of $\gamma$.  Because $\gamma$ enters the likelihood
function in the factor $e^{2 \gamma x}$, changing $\gamma$ has a
much larger impact on the likelihood of sites with many mutant
nucleotides than those with few. In other words, a single site with
a high frequency of mutant nucleotides is very strong evidence for
positive selection or low $|\gamma|$, whereas a site with a low
frequency of mutant nucleotides is not very strong evidence for the
opposite.  Thus even a very few sites at which multiple mutant
lineages are sampled, but are incorrectly assumed to be the result
of a single high-frequency mutant lineage by the original PRF
method, can cause large inaccuracies in the inferred $\gamma$.
These inaccuracies in $\gamma$ then force corresponding
inaccuracies in the inferred $\thetasite$.

Previous simulation studies have not observed these problems with the
PRF method, and they appear to show that the PRF method makes accurate
inferences of mutation rates and selection pressures
\cite{bustamante01}.  However, these simulations themselves implicitly
assume infinite sites, and hence they cannot be used to test this aspect
of the PRF method.  In this paper, we have simulated a finite number of
sites that evolve according to the Wright-Fisher model that forms the
basis of the PRF derivation. As our simulations and fits demonstrate,
the infinite-sites assumption causes the original PRF method to
systematically underestimate selective pressures and mutation rates, and
to find positive selection where it does not in fact exist.  The
techniques developed here, by contrast, produce accurate estimates of
mutation rates and selection pressures across a broad range of
biologically reasonable parameters.

The fact that a few highly polymorphic sites have a large impact on the
inferred values of $\gamma$ and $\thetasite$ points to another important
aspect of the original PRF method.  In assuming that all sites have the
same value of $\gamma$, the PRF method infers some sort of ``weighted
average'' of the variable selection strengths across sites.  This is not
necessarily a bad thing --- in the face of limited data, inferring the
full distribution of selection strengths is impossible, and we want
instead to have a rough sense of the average strength and direction of
selection.  However, the exact nature of this ``weighting'' is unclear.
Since a single highly polymorphic site has a much larger effect in
reducing the inferred value of $|\gamma|$ (or in suggesting positive
selection) than a site with a low frequency of mutant nucleotides has in
doing the reverse, the weighting clearly emphasizes neutral, nearly
neutral, or positively selected sites more than deleterious sites.  In
particular, since the PRF method ignores monomorphic sites, it does not
weight sites at which mutations are lethal.  Thus, the weighting of the
original PRF method is useful for increasing the sensitivity of the
method to detecting positive selection, but the details of how this
weighting works are poorly understood despite being crucial for
understanding what an inference of positive selection really means.

The methods we have developed allow us to relax the assumption that
all sites have the same $\gamma$, and instead infer aspects of the
distribution of selection pressures across sites. It is not yet
clear how much data is required to provide adequate power for
inferring this distribution to a given resolution. However, we can
hope to gain a great deal of insight with only a few additional
parameters --- say, the number of sites which are neutral, lethal,
and negatively selected, and the ``weighted average'' selection
pressure on the latter class. As we have shown, we can estimate the
number of lethal sites with no reduction in power relative to the
original PRF method, so this proposal would only involve one
additional parameter.  Additional classes of sites would involve
one or two extra parameters each, depending on whether we specify
or infer the selection pressure operating on these sites.  The
appropriate choice of resolution will depend on the context and
quality of the data.  We hope that experience gained from these
approaches will also shed light on the nature of the weighting when
a single ``average'' selection pressure is inferred from the same
data, and allow us to interpret this number more precisely.

Unlike the original PRF method, our two diffusion methods do not
offer an easy way to infer the existence of positive selection.
This limitation arises because positive selection is inherently out
of equilibrium.  The original PRF method and our modified, per-site
PRF method handle this issue by positing rather strange dynamics at
individual sites.  All mutant lineages are assumed to be positively
selected --- so if a mutant nucleotide fixes at a given site,
mutations back to the ancestral nucleotide are again assumed to be
positively selected.  While this assumption makes little sense at a
per-site level, it allows us to obtain a steady state across sites,
provided positive selection is ongoing and not saturated.  We could
modify our diffusion methods to mimic the PRF treatment of positive
selection by changing the boundary conditions in our diffusion
equations.  Specifically, we would assume that probability flowing
into $x=1$ (i.e.  fixation of a mutant nucleotide) is absorbed and
moved to $x=0$ (i.e.  ``reset'' so that new mutations will again be
favored).  This diffusion equation can be solved exactly, and the
solution used as a basis for inferring positive selection using the
the per-site diffusion methods we have developed.

Ideally, however, we would like to infer positive selection in the
context of a realistic and well-defined model of the dynamics at
individual sites.  Such an approach would necessarily involve
solutions for the transient dynamics of positively selected
mutations sweeping through a population. \citet{kimura55a} found
the full time-dependent solution for the diffusion equation
describing the dynamics of a positively selected allele. Such
processes are initiated as mutations arise, at Poisson-distributed
times. Thus we could construct an expected frequency distribution
across sites consisting of a superposition of time-dependent
solutions at each site, and use this as the basis for inference of
positive selection.  In this paper, we do not pursue either of
these methods for detecting positive selection within the per-site
diffusion framework, but this remains an important direction for
future work.

Regardless of the methodology, it will always be difficult to
discriminate positive selection from negative selection on the basis of
the polymorphism frequency spectrum alone, particularly when only folded
data are available.  Whether selection is positive or negative, mutant
lineages drift nearly neutrally when their frequency is between $0$ and
$\frac{1}{|\gamma|}$.  Positively selected lineages then fix relatively
quickly once their frequency becomes substantially larger than
$\frac{1}{|\gamma|}$, while negatively selected lineages rarely ever
reach frequencies greater than $\frac{1}{|\gamma|}$.  Thus from the
point of view of the polymorphism frequency spectrum, positive selection
is similar to random drift on $\left[ 0, 1/|\gamma| \right]$, with the
upper bound a roughly absorbing boundary condition.  Negative selection,
on the other hand, is also similar to random drift on $\left[ 0,
1/|\gamma| \right]$, but with the upper bound a roughly reflecting
boundary condition.  Although this is relatively crude --- selection
does in fact have some impact on low-frequency lineages, and the
boundary conditions are not exactly absorbing or reflecting --- it
indicates that the polymorphism frequency spectrum is roughly similar
for negative and positive selection at the same $|\gamma|$.  Thus power
to distinguish positive from negative selection based on the
polymorphism frequency spectrum, especially with folded data, will
always be relatively limited, regardless of the method used.

Inference methods that utilize both interspecies divergence as well
as intraspecies polymorphism at synonymous and nonsynonymous sites
(i.e. McDonald-Kreitman type tests \cite{mcdonaldkreitman91}) are
often superior to those that rely on the full intraspecific
polymorphism frequency spectrum.  While these methods lose some
power by ignoring information about the full polymorphism frequency
spectrum, they utilize data from more than one species as well as a
key biological assumption -- that synonymous sites are neutral --
to provide a sort of internal control. As a result, such methods
are typically less sensitive to many of the assumptions of the PRF
model \cite{loewe06, sawyerhartl92}, including the infinite-sites
assumption. Compared to the original PRF method, violation of the
infinite-sites assumption causes less severe errors under
McDonald-Kreitman type inferences, because highly polymorphic sites
no longer have a disproportionate impact on inferred parameters.
The quantification of such biases and the development of a
finite-site  framework for McDonald-Kreitman type inferences remain
topics for future research.

\clearpage

\newpage

\bibliographystyle{genetics}
\bibliography{prfbib}

\begin{thebibliography}{29}

\bibitem[{\sc Akashi}(1995)]{akashi95}
{\sc Akashi, H.}, 1995 Inferring weak selection from patterns of polymorphism
  and divergence at "silent" sites in drosophila dna. Genetics {\bf 139}:
  1067--1076.

\bibitem[{\sc Akashi}(1999)]{akashi99}
{\sc Akashi, H.}, 1999 Inferring the fitness effects of dna mutations from
  polymorphism and divergence data: Statistical power to detect directional
  selection under stationarity and free recombination. Genetics {\bf 151}:
  221--238.

\bibitem[{\sc Akashi {\rm and} Schaeffer}(1997)]{akashishaeffer97}
{\sc Akashi, H. {\rm and} S.~W. Schaeffer}, 1997 Natural selection and the
  frequency distributions of "silent" dna polymorphism in drosophila. Genetics
  {\bf 146}: 295--307.

\bibitem[{\sc Bartolome {\em et~al.\/}}(2005){\sc Bartolome, Maside, Yi, Grant,
  {\rm and} Charlesworth}]{bartolome05}
{\sc Bartolome, C., X.~Maside, S.~Yi, A.~L. Grant, {\rm and} B.~Charlesworth},
  2005 Patterns of selection on synonymous and nonsynonymous variants in
  drosophila miranda. Genetics {\bf 169}: 1495--1507.

\bibitem[{\sc Bustamante {\em et~al.\/}}(2005){\sc Bustamante, Fledel-Alon,
  Williamson, Nielsen, Hubisz, Glanowki, Tanenbaum, White, Sninsky, Hernandez,
  Civello, Adams, Cargill, {\rm and} Clark}]{bustamante05}
{\sc Bustamante, C.~D., A.~Fledel-Alon, S.~Williamson, R.~Nielsen, M.~T.
  Hubisz, S.~Glanowki, D.~M. Tanenbaum, T.~J. White, J.~J. Sninsky, R.~D.
  Hernandez, D.~Civello, M.~D. Adams, M.~Cargill, {\rm and} A.~G. Clark}, 2005
  Natural selection on protein-coding genes in the human genome. Nature {\bf
  437}: 1153--1157.

\bibitem[{\sc Bustamante {\em et~al.\/}}(2003){\sc Bustamante, Nielsen, {\rm
  and} Hartl}]{bustamantehartl03}
{\sc Bustamante, C.~D., R.~Nielsen, {\rm and} D.~L. Hartl}, 2003 Maximum
  likelihood and bayesian methods for estimating the distribution of selective
  effects among classes of mutations using dna polymorphism data. Theoretical
  Population Biology {\bf 63}: 91--103.

\bibitem[{\sc Bustamante {\em et~al.\/}}(2002){\sc Bustamante, Nielsen, Sawyer,
  Olsen, Purugganan, {\rm and} Hartl}]{bustamante02}
{\sc Bustamante, C.~D., R.~Nielsen, S.~Sawyer, K.~M. Olsen, M.~D. Purugganan,
  {\rm and} D.~L. Hartl}, 2002 The cost of inbreeding in arabidopsis. Nature
  {\bf 416}: 531--534.

\bibitem[{\sc Bustamante {\em et~al.\/}}(2001){\sc Bustamante, Wakeley, Sawyer,
  {\rm and} Hartl}]{bustamante01}
{\sc Bustamante, C.~D., J.~Wakeley, S.~Sawyer, {\rm and} D.~L. Hartl}, 2001
  Directional selection and the site-frequency spectrum. Genetics {\bf 159}:
  1779--1788.

\bibitem[{\sc Ewens}(2004)]{ewensbook}
{\sc Ewens, W.~J.}, 2004 {\em Mathematical Population Genetics: I. Theoretical
  Introduction\/}. Springer, New York, NY.

\bibitem[{\sc Galtier {\em et~al.\/}}(2006){\sc Galtier, Bazin, {\rm and}
  Bierne}]{galtier06}
{\sc Galtier, N., E.~Bazin, {\rm and} N.~Bierne}, 2006 Gc-biased segregation of
  noncoding polymorphisms in drosophila. Genetics {\bf 172}: 221--228.

\bibitem[{\sc Hartl {\em et~al.\/}}(1994){\sc Hartl, Moriyama, {\rm and}
  Sawyer}]{hartlsawyer94}
{\sc Hartl, D.~L., E.~N. Moriyama, {\rm and} S.~A. Sawyer}, 1994 Selection
  intensity for codon bias. Genetics {\bf 138}: 227--234.

\bibitem[{\sc Kimura}(1955)]{kimura55a}
{\sc Kimura, M.}, 1955 Solution of a process of random genetic drift with a
  continuous model. PNAS {\bf 41}: 144--150.

\bibitem[{\sc Lercher {\em et~al.\/}}(2002){\sc Lercher, Smith, Eyre-Walker,
  {\rm and} Hurst}]{lercherhurst02}
{\sc Lercher, M.~J., N.~G.~C. Smith, A.~Eyre-Walker, {\rm and} L.~Hurst}, 2002
  The evolution of isochores: Evidence from snp frequency distributions.
  Genetics {\bf 162}: 1805--1810.

\bibitem[{\sc Loewe {\em et~al.\/}}(2006){\sc Loewe, Charlesworth, Bartolome,
  {\rm and} Noel}]{loewe06}
{\sc Loewe, L., B.~Charlesworth, C.~Bartolome, {\rm and} V.~Noel}, 2006
  Estimating selection on nonsynonymous mutations. Genetics {\bf 172}:
  1079--1092.

\bibitem[{\sc McDonald {\rm and} Kreitman}(1991)]{mcdonaldkreitman91}
{\sc McDonald, J. {\rm and} M.~Kreitman}, 1991 Adaptive protein evolution at
  the adh locus in drosophila. Nature {\bf 351}: 652--654.

\bibitem[{\sc Moran}(1959)]{moran59}
{\sc Moran, P. A.~P.}, 1959 The survival of a mutant gene under selection. ii.
  Journal of the Austrian Mathematical Society {\bf 1}: 485--491.

\bibitem[{\sc Nachman}(1998)]{nachman98}
{\sc Nachman, M.~W.}, 1998 Deleterious mutations in animal mitochondrial dna.
  Genetica {\bf 102/103}: 61--69.

\bibitem[{\sc Piganeau {\rm and} Eyre-Walker}(2003)]{piganeaueyrewalker03}
{\sc Piganeau, G. {\rm and} A.~Eyre-Walker}, 2003 Estimating the distribution
  of fitness effects from dna sequence data: Implications for the molecular
  clock. Proceedings of the National Academy of Sciences {\bf 100}:
  10335--10340.

\bibitem[{\sc Rand {\rm and} Kann}(1998)]{randkann98}
{\sc Rand, D.~M. {\rm and} L.~M. Kann}, 1998 Mutation and selection at silent
  and replacement sites in the evolution of animal mitochondrial dna. Genetica
  {\bf 102/103}: 393--407.

\bibitem[{\sc Sawyer {\em et~al.\/}}(2003){\sc Sawyer, Kulathinal, Bustamante,
  {\rm and} Hartl}]{sawyerhartl03}
{\sc Sawyer, S., R.~J. Kulathinal, C.~D. Bustamante, {\rm and} D.~L. Hartl},
  2003 Bayesian analysis suggests that most amino acid replacements in
  drosophila are driven by positive selection. Journal of Molecular Evolution
  {\bf 57}: S154--S164.

\bibitem[{\sc Sawyer {\rm and} Hartl}(1992)]{sawyerhartl92}
{\sc Sawyer, S.~A. {\rm and} D.~L. Hartl}, 1992 Population-genetics of
  polymorphism and divergence. Genetics {\bf 132}: 1161--1176.

\bibitem[{\sc Wakeley}(2003)]{wakeley03}
{\sc Wakeley, J.}, 2003 Polymorphism and divergence for island-model species.
  Genetics {\bf 163}: 411--420.

\bibitem[{\sc Watterson}(1977)]{watterson77}
{\sc Watterson, G.~A.}, 1977 Heterosis or neutrality? Genetics {\bf 85}:
  789--814.

\bibitem[{\sc Weinreich {\rm and} Rand}(2000)]{weinreichrand00}
{\sc Weinreich, D.~M. {\rm and} D.~M. Rand}, 2000 Contrasting patterns of
  nonneutral evolution in proteins encoded in nuclear and mitochondrial
  genomes. Genetics {\bf 156}: 385--399.

\bibitem[{\sc Williamson {\em et~al.\/}}(2004){\sc Williamson, Fledel-Alon,
  {\rm and} Bustamante}]{williamson04}
{\sc Williamson, S., A.~Fledel-Alon, {\rm and} C.~D. Bustamante}, 2004
  Population genetics of polymorphism and divergence for diploid selection
  models with arbitrary dominance. Genetics {\bf 168}: 468--475.

\bibitem[{\sc Williamson {\em et~al.\/}}(2005){\sc Williamson, Hernandez,
  Fledel-Alon, Zhu, Nielsen, {\rm and} Bustamante}]{williamson05}
{\sc Williamson, S.~H., R.~Hernandez, A.~Fledel-Alon, L.~Zhu, R.~Nielsen, {\rm
  and} C.~D. Bustamante}, 2005 Simultaneous inference of selection and
  population growth from patterns of variation in the human genome. Proceedings
  of the National Academy of Sciences {\bf 102}: 7882--7887.

\bibitem[{\sc Wright}(1938)]{wright38}
{\sc Wright, S.}, 1938 The distribution of gene frequencies under irreversible
  mutation. Proceedings of the National Academy of Sciences {\bf 24}: 253--259.

\bibitem[{\sc Wright}(1949)]{wright49}
{\sc Wright, S.}, 1949 Adaptation and selection. In {\em Genetics,
  Paleontology, and Evolution\/}, edited by G.~L. Jepson, G.~G. Simpson, {\rm
  and} E.~Mayr, pp. 365--389, Princeton University Press, Princeton, NJ.

\bibitem[{\sc Zhu {\rm and} Bustamante}(2005)]{zhubustamante05}
{\sc Zhu, L. {\rm and} C.~D. Bustamante}, 2005 A composite-likelihood approach
  for detecting directional selection from dna sequence data. Genetics {\bf
  170}: 1411--1421.

\end{thebibliography}

\clearpage
\newpage

\begin{figure}[ht] \begin{center}
\epsfig{file=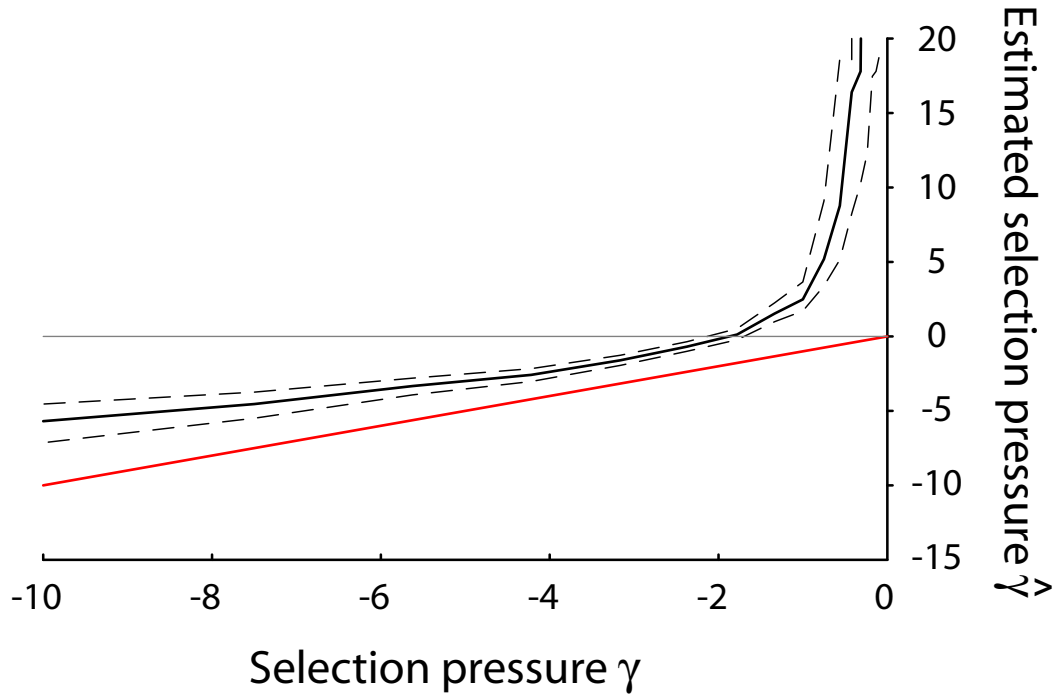,angle=0,width=14cm}
\caption{Maximum-likelihood estimates of selection pressures
obtained under the PRF method for unfolded polymorphism data. The
figure shows the estimated selection pressure $\hat{\gamma}$, on
the $y$-axis, obtained by applying the PRF method to $n=14$
sequences sampled from a simulated Wright-Fisher population (see
Accuracy of Inference Techniques).  Dashed lines indicate the 95\%
confidence interval around the ML estimate.  The true selection
pressure $\gamma$ used in the simulations is shown on the $x$-axis.
The line $\hat{\gamma}=\gamma$ (i.e. a perfect prediction) is shown
in red, and the line $\hat{\gamma}=0$ is shown in gray.  The
mutation rate is $\theta_s=0.5$ per site.  The PRF method
systematically underestimates the strength of selection, and it
often leads to erroneous inferences of strong positive selection
when selection is negative. In all cases, the PRF method strongly
rejects the true parameters (the difference in the natural log
likelihood between $\hat{\gamma}$ and $\gamma$ ranges from 14 to
over 700).} \label{TritEquil} \end{center} \end{figure}

\clearpage
\newpage

\begin{figure}[ht] \begin{center}
\epsfig{file=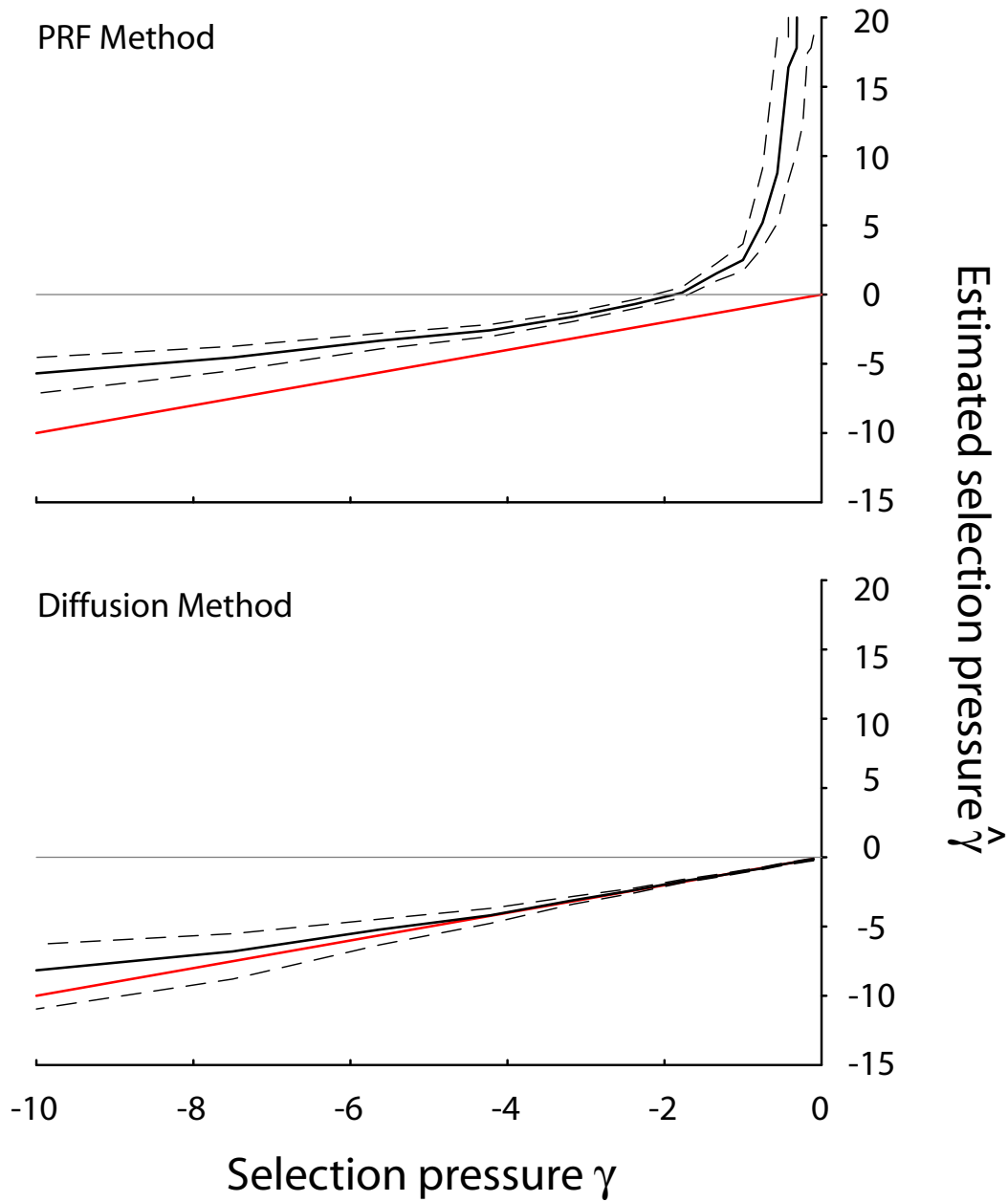,angle=0,width=14cm} \caption{A
comparison between the accuracy of inferred selection pressures
under the PRF method (top, as in Fig. 1) versus the one-dimensional
diffusion method (bottom). In both cases, selection pressures were
estimated from the unfolded polymorphism frequencies among $n=14$
sequences sampled from a simulated Wright-Fisher population. Dashed
lines indicate 95\% confidence intervals around the ML estimates.
The line $\hat{\gamma}=\gamma$ is shown in red, and the line
$\hat{\gamma}=0$ is shown in gray.  Unlike the PRF method, the
diffusion method provides unbiased estimates of the selection
pressure, and it does not lead to erroneous inferences of positive
selection.} \label{figure2}
\end{center} \end{figure}

\clearpage
\newpage

\begin{figure}[ht] \begin{center}
\epsfig{file=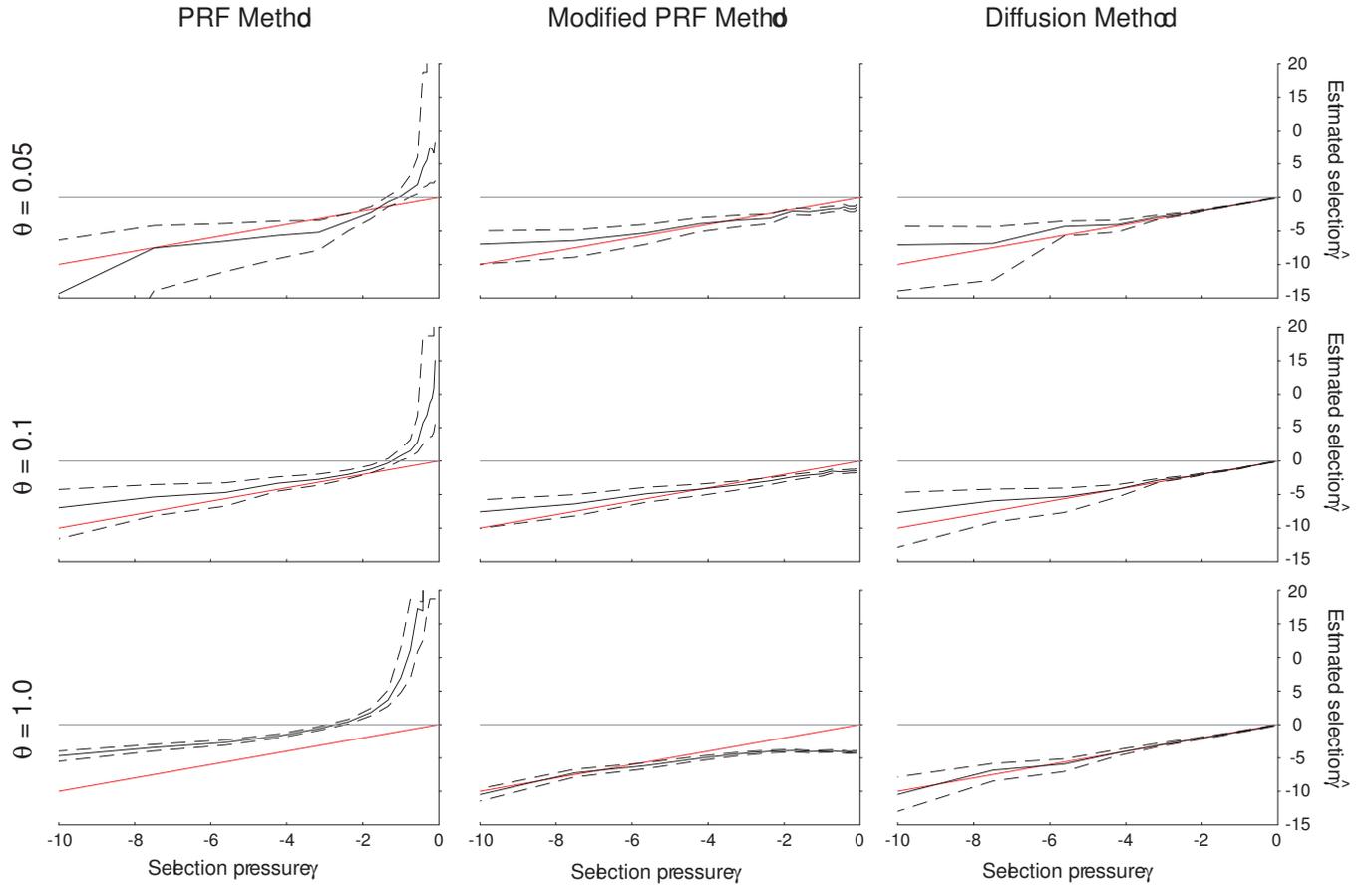,angle=0,width=18cm}
\caption{Maximum-likelihood estimates of selection pressures
obtained under the PRF method, the modified (per-site) PRF method,
and the one-dimensional diffusion method. Selection pressures were
estimated from the unfolded polymorphism frequencies among $n=14$
sequences sampled from a simulated Wright-Fisher population. In
each panel, the simulated selection pressure $\gamma$ is shown on
the $x$-axis, and the estimated selection pressure $\hat{\gamma}$
on the $y$-axis. Dashed lines indicate 95\% confidence intervals
around the ML estimates. The line $\hat{\gamma}=\gamma$ is shown in
red, and the line $\hat{\gamma}=0$ is shown in gray.  Simulations
and fits were performed across a range of mutation rates, shown in
separate rows. The diffusion method and, to a lesser extent, the
modified PRF method correct the biases inherent in the original PRF
method, especially at large mutation rates.} \label{figure3}
\end{center}
\end{figure}

\clearpage
\newpage

\begin{figure}[ht] \begin{center}
\epsfig{file=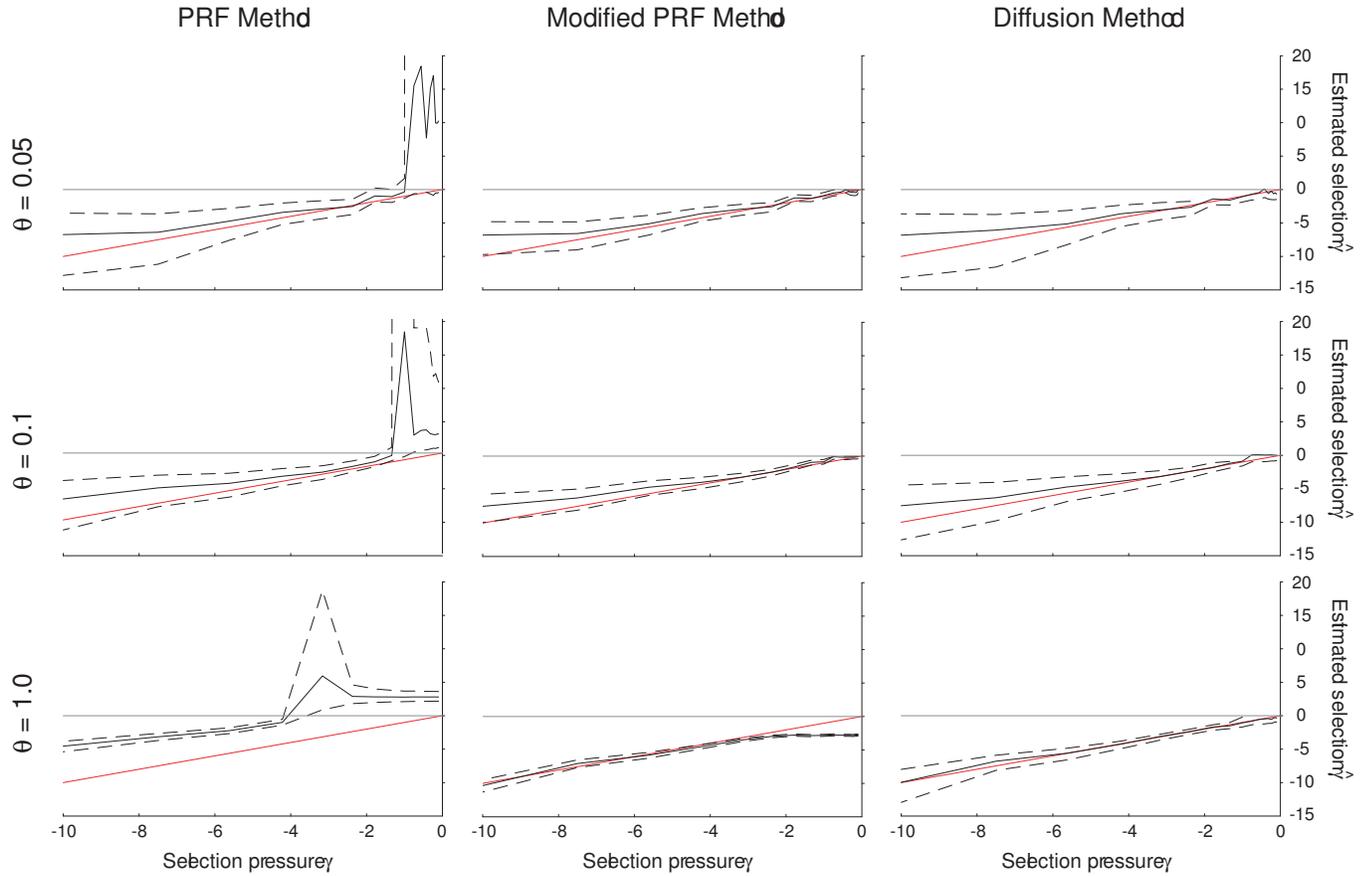,angle=0,width=18cm} \caption{
Maximum-likelihood estimates of selection pressures obtained under
the PRF method, the modified (per-site) PRF method, and the
three-dimensional diffusion method. Selection pressures were
estimated from the folded polymorphism frequencies among $n=14$
sequences sampled from a simulated Wright-Fisher population. The
diffusion method and, to a lesser extent, the modified PRF method
correct the biases inherent in the original PRF method. When
selection is weakly negative, the diffusion method cannot reject
positive selection on the basis of folded data, as indicated by the
lack of the upper confidence interval for some of the fits.}
\end{center} \end{figure}

\clearpage
\newpage

\begin{figure}[ht] \begin{center}
\epsfig{file=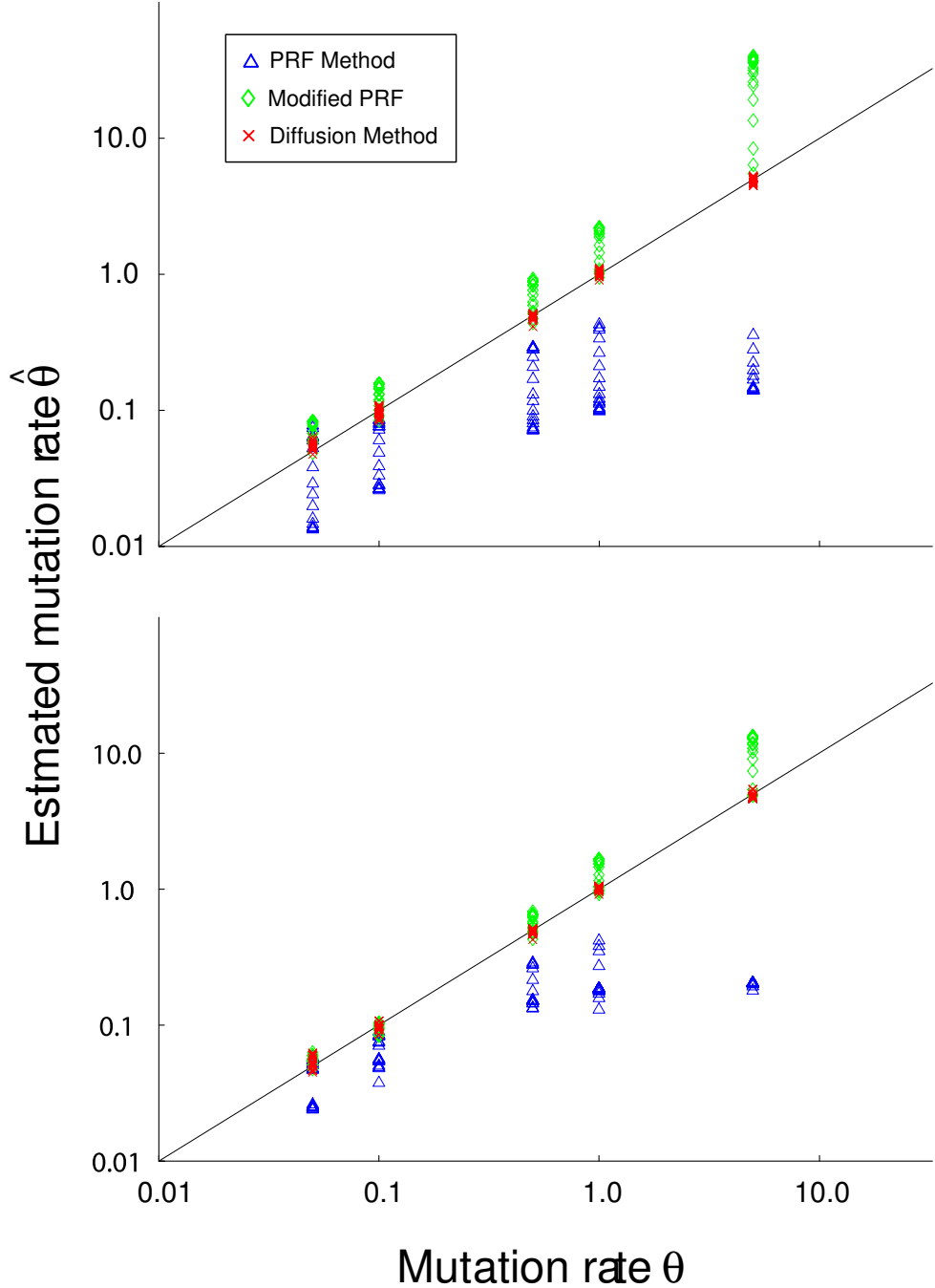,angle=0,width=13cm}
\caption{Maximum-likelihood estimates of the mutation rate,
$\thetasite = 2 N \mu$, obtained under the PRF method (blue), the
modified PRF method (green), and the diffusion methods (red).
Mutation rates were estimated from the unfolded (top) and folded
(bottom) polymorphism frequencies among $n=14$ sequences sampled
from a simulated Wright-Fisher population. The simulated mutation
rate $\thetasite$ is shown on the $x$-axis, and the estimated
mutation rates $\hat{\thetasite}$ on the $y$-axis. The line
$\hat{\thetasite}=\thetasite$ is shown in black.  For each value of
$\thetasite$, simulations and fits are shown for 17 different
values of $\gamma$, ranging from $\gamma=-10.0$ to $\gamma=-0.1$.
The PRF method systematically underestimates the mutation rate,
especially when selection is weak. The diffusion methods provide
accurate and unbiased estimates of the mutation rate for both
folded and unfolded data, across the full range of parameters.}
\end{center} \end{figure}

\clearpage
\newpage

\begin{table}
\begin{center}
\begin{tabular}{|c|c|c|}
\hline
\multicolumn{3}{|c|}{Accuracy of Predicted Number}\\
\multicolumn{3}{|c|}{of Monomorphic Sites}\\
\hline
Actual $\thetasite$ & Average Error & Median Error \\
\hline
$0.05$ & $7.6 \%$ & $ 7.8 \%$\\
$0.1$ & $3.3 \%$ & $ 1.8 \%$\\
$0.5$ & $2.3 \%$ & $ 1.1 \%$\\
$1.0$ & $1.4 \%$ & $ 0.5 \%$\\
$5.0$ & $0.1 \%$ & $ 0.1 \%$\\
\hline
\end{tabular}
\end{center}
\caption{\label{table1} 
Accuracy of estimates for the expected number of monomorphic sites.
We calculated
maximum-likelihood estimates of $\gamma$ and $\thetasite$
using the one-dimensional diffusion method applied to
unfolded simulated data excluding monomorphic sites. From
these values, we calculated the expected number of monomorphic sites.
Shown are the differences between the observed and expected number of
monomorphic sites (in a simulated gene of length $L = 1000$ sites).  For
each value of $\thetasite$, we show both the average and median
differences in numbers of monomorphic sites, across simulations with
$\gamma$ ranging from $-0.1$ to $-10$. These results imply that we can
accurately estimate the number of sites monomorphic due to drift.
Thus, if a gene contains a similar or larger number of lethal site, we
can
also estimate the number of such lethal sites to within the above
accuracy. 
}
\end{table}
\end{document}